\documentclass[aps,pre,showpacs,showkeys,amssymb,twocolumn]{revtex4}
\usepackage{bm}
\usepackage{graphicx}
\usepackage{amsfonts}
\usepackage{amsmath}
\usepackage{amssymb}
\usepackage{xcolor}

\begin{document}

\title{Necessary and sufficient conditions for $\mathbb{Z}_2$-symmetry-breaking phase transitions}
%Phase transitions triggered by dumbbell-shaped equipotential hypersurfaces

\author{Fabrizio Baroni}

\email{f.baroni@ifac.cnr.it, baronifab@libero.it}
\affiliation{IFAC-CNR Institute of applied physics "Nello Carrara", Via Madonna del Piano 10, I-50019 Sesto Fiorentino (FI), Italy}

\date{\today}

\begin{abstract}
	In a recent paper a toy model (\emph{hypercubic model}) undergoing a first-order $\mathbb{Z}_2$-symmetry-breaking phase transition ($\mathbb{Z}_2$-SBPT) was introduced. The hypercubic model was inspired by the \emph{topological hypothesis}, according to which a phase transition may be entailed by suitable topological changes of the equipotential surfaces ($\Sigma_v$'s) of configuration space. In this paper we show that at the origin of a $\mathbb{Z}_2$-SBPT there is a geometric property of the $\Sigma_v$'s, i.e., dumbbell-shaped $\Sigma_v$'s suitably defined, which includes a topological change as a limiting case. This property is necessary and sufficient condition to entail a $\mathbb{Z}_2$-SBPT. This new approach has been applied to three models: a modified version introduced here of the hypercubic model, a model introduced in a recent paper with a continuous $\mathbb{Z}_2$-SBPT belonging to several universality classes, and finally to a physical models, i.e., the mean-field $\phi^4$ model and a simplified version of it.
\end{abstract}



\maketitle

\section{Introduction}

Phase transition are very common in nature. They are sudden changes of the macroscopic behavior of a natural system composed by many interacting parts occurring while an external parameter is varied. Phase transitions are an example of emergent behavior, i.e., of a collective properties having no direct counterpart in the dynamics or structure of individual atoms \cite{lebowitz}.

From a statistical-mechanical viewpoint, in the canonical ensemble, describing a system at constant temperature $T$, a phase transition occurs at special values if the temperature called transition points, where thermodynamic quantities such as pressure, magnetization, or heat capacity, are non-analytic functions of $T$ and some system symmetries break spontaneously. These points are the boundaries between different phases of the system. Starting from the celebrated solution of the 2d Ising model by Onsager, i.e., an example of a $\mathbb{Z}_2$-symmetric system, and later developments like the renormalization group theory \cite{goldenfeld}, our knowledge of the properties of phase transitions have considerably deepened.

Yet, the situation is not completely satisfactory. First, in the canonical ensemble phase transitions occur only in the case of infinite systems: following an early suggestion by Kramers \cite{c}, Lee and Yang \cite{yl} showed that the thermodynamic limit $N\rightarrow\infty$ ($N$ is the number of degrees of freedom, and the limit is taken at fixed density) must be invoked to explain the existence of singularities in the partition function $Z(T)$ and then in the thermodynamic functions defined as derivatives of $Z(T)$. Since in the last decades many examples of transitional phenomena in systems far form the thermodynamic limit have been found (e.g., in nuclei, atomic clusters, biopolymers, superconductivity, superfluidity), a description of phase transition valid also for finite systems would be desirable. Second, while necessary conditions for the presence of a phase transitions can be found, not much is known about sufficient conditions: no general procedure is at hand to tell if a system where a phase transition is not ruled out from the beginning does have or not such a transition without computing $Z$. This might indicate that our understanding of this phenomenon is still incomplete. 

These considerations motivate a study of the nature of phase transitions which may be based on alternative approaches. One of such approaches, proposed in Ref. \cite{cccp} and developed later in Ref. \cite{ccp}, is based on concepts and tools drawn from differential geometry and topology. The main issue of this new approach is the \emph{topological hypothesis}, whose content is that phase transitions are due to a topology change of the equipotential surfaces $\Sigma_v$'s of configuration space, those where the system lives as the number of its degrees of freedom becomes very large. This idea has been discussed and tested in many papers \cite{acprz,b,bc,ccp1,ckn,gss,gm,k,rs,dgppdfv,pgfcp,pettini}. 

A first answer to the problem of what may be, if any, the sufficient conditions to entail phase transitions, either topological alone or in addition to geometric hypotheses, was given in Ref. \cite{bc}, where a straightforward theorem (Theorem 1 in the paper) for the occurrence of a $\mathbb{Z}_2$-symmetry-breaking phase transition ($\mathbb{Z}_2$-SBPT) was proven. 

In this paper we generalize that theorem finding out a more general sufficient, and even necessary, condition for $\mathbb{Z}_2$-SBPTs to occur. The sufficient condition of Theorem 1 in Ref. \cite{bc} makes use only of topological properties of the equipotential surfaces ($\Sigma_v$'s), while the generalized version is given in terms of a geometric property of the $\Sigma_v$'s largely independent on their topology. The $\Sigma_v$'s with this property have been called \emph{dumbbell-shaped}, meaning the presence of a narrow neck connecting two major lobes. The original topological sufficient condition survives as a limiting case. 

The paper is structured in a pedagogical way. In Sec. \ref{hcm} we start by revisiting a toy model (hypercubic model) undergoing a first-order $\mathbb{Z}_2$-SBPT introduced in Ref. \cite{bc}. This shows, in an intuitive way, how the framework of dumbbell-shaped $\Sigma_v$'s works in practice. In Sec.  \ref{strangled} we give the rigorous definition of dumbbell-shaped $\Sigma_v$, and give the necessary and sufficient condition for $\mathbb{Z}_2$-SBPTs. In Sec. \ref{bs} we introduce a modified version of the hypercubic model which rigorously shows what depicted in Sec. \ref{hcm} from a mathematical viewpoint. Finally, in Sec. \ref{phi4} we apply the framework of dumbbell-shaped $\Sigma_v$'s to a physical model, i.e., the mean-field $\phi^4$ model and to a simplified version of it  introduced in Ref. \cite{b_5}.

\section{The origin of the $\mathbb{Z}_2$-SB in the hypercubic model}
\label{hcm}

In this section we analyze the generating-mechanism of the $\mathbb{Z}_2$-SB in a toy model (called \emph{hypercubic model}) introduced in Ref. \cite{bc}. The model was defined in such a way to undergo a first-order $\mathbb{Z}_2$-SBPT. The potential is nothing but a generalization to $N$ dimensions of a square double-well potential with the gap between the wells proportional to the number of degrees of freedom $N$. The kinetic term is standard and is not taken into consideration here because it does not contribute to the $\mathbb{Z}_2$-SB. 

\begin{figure}
	\begin{center}
		\includegraphics[width=0.235\textwidth]{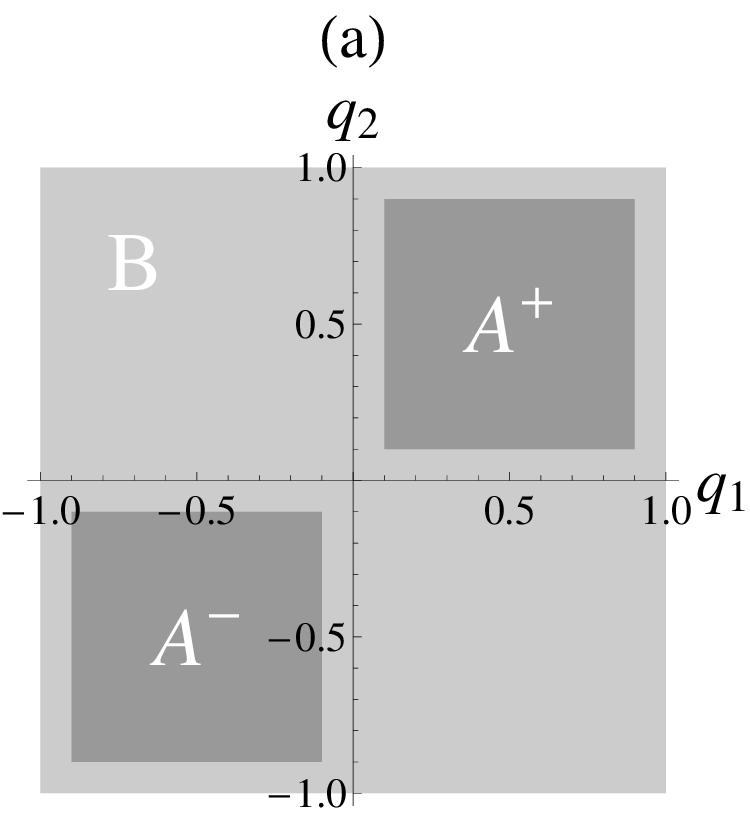}
		\includegraphics[width=0.235\textwidth]{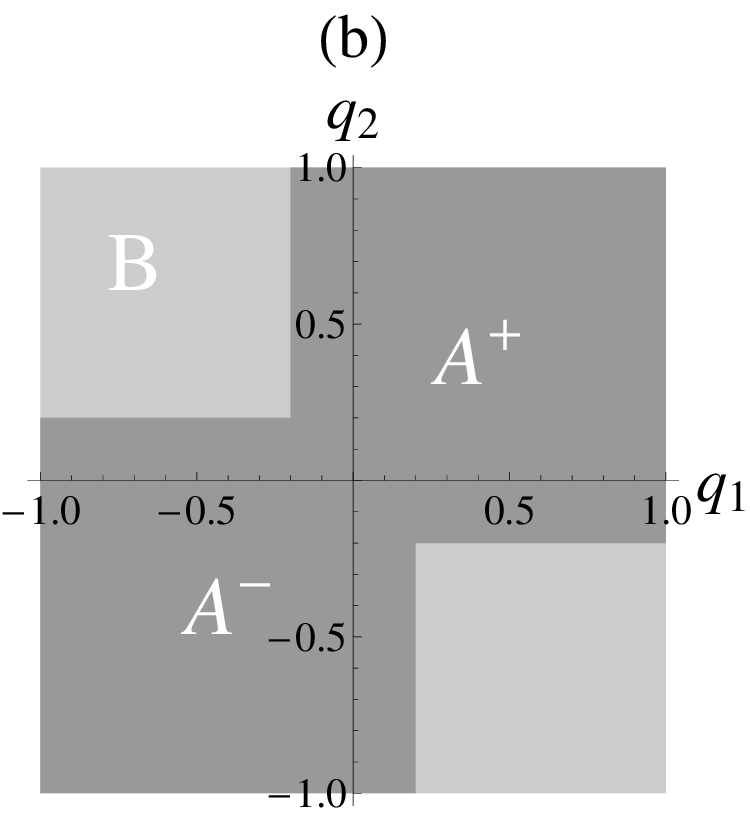}
		\caption{(a) Sketch of the $\Sigma_{v,2}$'s (\ref{Vhcm}) of the hypercubic model (\ref{Vhcm}) with $A^+\cap A^-=\emptyset$. (b) The same of panel (a) for $b=2$ and $a=1.2$ in such a way to have $A^+\cap A^-\neq\emptyset$.}
		\label{cubes}
	\end{center}
\end{figure}

The potential is defined in terms of equipotential surfaces $\Sigma_{v,N}$'s of configuration space $M\subseteq\mathbb{R}^N$ defined as follows
\begin{equation}
\Sigma_{v,N}=\{\mathbf{q}\in M: v(\mathbf{q})=v\},
\label{sigmavN}
\end{equation}
where $v=V/N$ is the specific potential or potential density. The $\Sigma_{v,N}$'s of the hypercubic model were defined as follows
\begin{equation}
\Sigma_{v,N}=\begin{cases}
\emptyset& \text{ if }\quad v<-v_c 
\\ 
A^+\cup A^-& \text{ if }\quad v=-v_c 
\\ 
\emptyset& \text{ if }\quad -v_c<v<0
\\ 
B\backslash(A^+\cup A^-)& \text{ if }\quad v=0 
\\ 
\emptyset& \text{ if }\quad v>0
\end{cases},
\label{Vhcm}
\end{equation}
where $v_c>0$, $B$ is an $N$-cube of side $b$ centered in the origin of coordinates, and $A^+, A^-$ are $N$-cubes of side $a$ disposed in such a way to be included in $B$ and to be one the image of the other under the $\mathbb{Z}_2$ symmetry (see Fig. \ref{cubes}). Furthermore, $b\geq 2a$ was assumed in such a way to have $A^+\cap A^-=\emptyset$. A topological change occurs in the $\Sigma_{v,N}$'s as the potential jumps between $-v_c$ and $0$. 

\begin{figure}
	\begin{center}
		\includegraphics[width=0.235\textwidth]{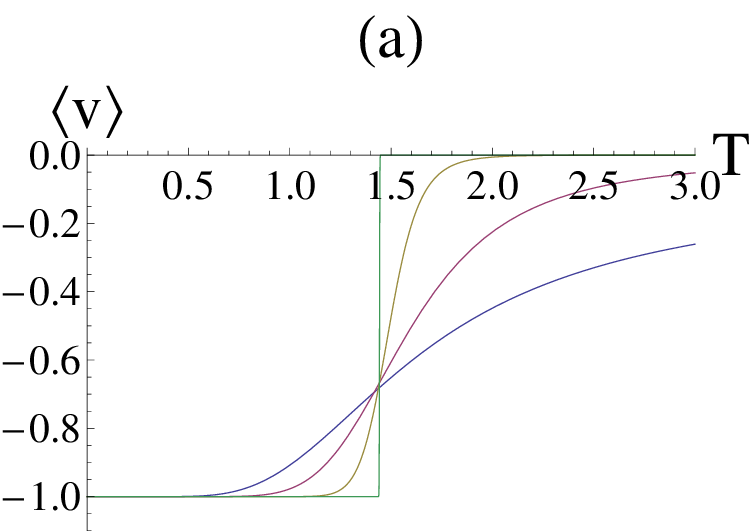}
		\includegraphics[width=0.235\textwidth]{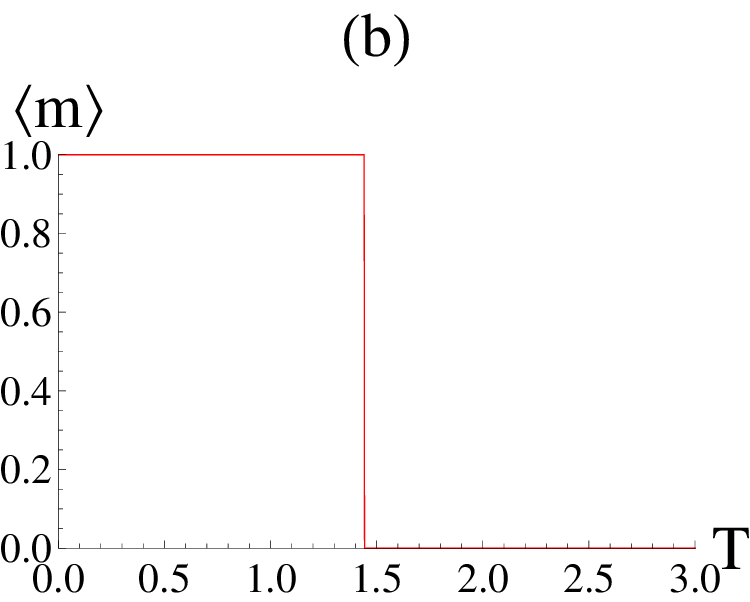}
		\caption{Hypercubic model (\ref{Vhcm}) with $v_c=1$ $a=1.5$, and $b=2$. (a) Specific average potential $\left\langle v\right\rangle$ as a function of the temperature $T$. The smooth lines are for $N=30, 50, 100$ (green, blue, magenta), they show a non-uniform convergence toward a discontinuous limit for $N\to\infty$ (green line) corresponding to a first-order PT with critical temperature $T_c=1/\ln(4/3)$. (b) The same of panel (a) for the spontaneous magnetization $\left\langle m\right\rangle$.}
		\label{termohcm}
	\end{center}
\end{figure}

The canonical thermodynamic can be analytically solved in a very simple way. For our purposes we report only the specific average potential and the spontaneous magnetization (see Fig. \ref{termohcm})
\begin{equation}
\left\langle v\right\rangle=\left\{\begin{array}{ll}
-v_c & \hbox{if} \quad T<T_c
\\
0 & \hbox{if} \quad T>T_c
\end{array}\right.,
\end{equation}
\begin{equation}
\left\langle m\right\rangle=\left\{\begin{array}{ll}
\pm\frac{b-a}{2} & \hbox{if} \quad T<T_c
\\
0 & \hbox{if} \quad T>T_c.
\end{array}\right.,
\end{equation}
where $T_c=v_c/\ln(b/a)$ is the critical temperature. The Boltzmann constant $k$ has been assumed equal to $1$. The magnetization in the broken phase was assumed to be the center of mass coordinates of the $N$-cubes $A^+$, or $A^-$, because the ergodic hypothesis was assumed. Summarizing, the hypercubic model shows the complete picture of a first-order $\mathbb{Z}_2$-SBPT.

In Ref. \cite{bc} the authors assumed $b\ge 2a$ because they were guided by the idea that, in order to entail the $\mathbb{Z}_2$-SBPT, at least a topological change in the $\Sigma_{v,N}$'s were needed. The idea was that, if the $N$-cubes $A^+$, $A^-$ are disjointed, the probability of the representative point (RP) to jump between $A^+$, and $A^-$ is vanishing in the thermodynamic limit. Here, we show that the disjointedness of $A^+$, $A^-$ is not a necessary hypothesis, and the $\mathbb{Z}_2$-SBPT can be entailed by a weaker condition on $A^+\cup A^-$ which includes the disjointedness as a limiting case.

We start by observing that the solution of the thermodynamic does not require the restriction $b\ge 2a$, and it makes sense also under the less restrictive condition $b>a$. Indeed, $T_c=v_c/\ln(b/a)>0$ needs only $b>a$, and if $b=a$, then $T_c=0$ and the $\mathbb{Z}_2$-SBPT disappears. But, as $b>a>b/2$, $A^+\cap A^-\neq\emptyset$. Thus we observe: how can the $\mathbb{Z}_2$ symmetry be broken as $T<T_c$? And, what happens to the spontaneous magnetization?  Assuming the 'a priori' equal probability hypothesis, i.e., the ergodic hypothesis, the spontaneous magnetization should be vanishing because $A^+\cup A^-$ is $\mathbb{Z}_2$-symmetric. 

To help intuition, consider the case in which $a$ is only slightly greater than $b/2$. $A^+\cap A^-$ is a small $N$-cube of side $2a-b$ (see Fig. \ref{cubes}). How can the RP go freely around the whole $A^+\cup A^-$ if it has to pass through $A^+\cap A^-$? It is clear that the $\mathbb{Z}_2$ symmetry has to break despite that $A^+\cap A^-\neq\emptyset$. Actually, it is not important how small $A^+\cap A^-$ is. Indeed, the ratio between the volume of $A^+\cap A^-$, i.e., $(2a-b)^N$, and the volume of $A^+$ or $A^-$, i.e., $a^N$, is vanishing in the limit $N\to\infty$ because $2a-b<a$. This ratio is the probability to find the RP just in $A^+\cap A^-$ instead of $A^+$ or $A^-$, so that $A^+\cup A^-$ becomes a barrier that the RP cannot overcome. Summarizing, the crucial feature of $A^+\cup A^-$ that breaks the $\mathbb{Z}_2$ symmetry is to be "dumbbell-shaped", or in other words, we can say that $A^+\cup A^-$ has a "neck". This crucial concept is rigorously defined in Sec. \ref{strangled}.

More precisely, we are in front of the ergodicity breaking phenomenon, that, among its consequences, includes spontaneous SB. We recall that the ergodic hypothesis implies that we can substitute the time average of an observable $F(\mathbf{q})$ of the system, i.e., a function of the coordinates, by the canonical (or microcanonical) ensemble average
\begin{align}
\begin{split}
\overline{F}&=\lim_{t\rightarrow+\infty}\frac{1}{t}\int^{t}_{0}dt \,F(\mathbf{q}(t))
\\
&=\left\langle F\right\rangle=\frac{1}{Z(\beta)}\int_M d\mathbf{q} \,e^{-\beta V(\mathbf{q})} F(\mathbf{q}),
\end{split}
\end{align}
where $\beta=1/T$. In the case in which $M$, or a subset of it where the RP is forced to lay, has a neck, it is no longer possible to apply the formula above for the reasons aforementioned. The situation is in somewhat extent paradoxical. The assumption of the ergodic hypothesis leads to the consequence that it itself cannot be applied any more because of the particular shape of the $\Sigma_{v,N}$. We think that this is the true origin of all spontaneous SB phenomena, not only for a $\mathbb{Z}_2$-symmetric system.

\section{Dumbbell-shaped $\Sigma_{v,N}$'s}
\label{strangled}

Consider an $N$ degrees of freedom Hamiltonian system with a $\mathbb{Z}_2$ symmetry. Let us define the hyperplane of $\mathbb{R}^N$ at constant magnetization
\begin{equation}
\Sigma_{m,N}=\{\mathbf{q}\in\mathbb{R}^N: \frac{1}{N}\sum^{N}_{i=1}q_i=m\}.
\label{pim}
\end{equation}
The microcanonical volume, or density of states, at fixed $v$ and $m$ is
\begin{equation}
\omega_N(v,m)=\mu(\Sigma_{v,N}\cap\Sigma_{m,N})=\int_{\Sigma_{v,N}\cap\Sigma_{m,N}}\frac{d\Sigma}{\|\nabla V \land \nabla M\|},
\label{omegaNvm}
\end{equation}
where $\nabla V \land \nabla M$ is the Gram matrix, and $\Sigma_{v,N}$ is the equipotential surface defined in (\ref{sigmavN}).
$\omega_N(v,m)$ is linked to the microcanonical entropy $s_N(v,m)$ by the relation 
\begin{equation}
s_N(v,m)=\frac{1}{N}\ln \omega_N(v,m).
\end{equation}

\medskip
\noindent{\textbf{Definition 1.}} \emph{A $\Sigma_{v,N}$ is \emph{dumbbell-shaped} if $s_N(v,m)$ does not take the global maximum at $m=0$.}

\medskip
Knowing where the global maximum of $s_N(v,m)$ is located is crucial to determine the spontaneous magnetization $\left\langle m\right\rangle$. Indeed, the latter takes the $m$-value at which $s_N(v,m)$ is maximum. Let us see why. Consider the co-area formula \cite{fr} of the canonical partition function
\begin{equation}
Z_N=N\int_{v_{min}}^{+\infty}dv \,e^{-\beta Nv}\mu\left(\Sigma_{v,N}\right),
\end{equation}
where $v_{min}$ is the global minimum of the potential density $v$, which is assumed to be bounded from below. Since $\Sigma_{v,N}=\cup_{m\in\mathbb{R}}\left(\Sigma_{v,N}\cap\Sigma_{m,N}\right)$
\begin{eqnarray}
\mu\left(\Sigma_{v,N}\right)=\sqrt{N}\int dm\,\mu(\Sigma_{m,N}\cap\Sigma_{m,N})\nonumber
\\
=\sqrt{N}\int dm\,e^{N s_N(v,m)},
\end{eqnarray}
where $\sqrt{N}$ arises because for an increment $dm$ of $m$ the distance between $\Sigma_{m,N}$ and $\Sigma_{m+dm,N}$ is $\sqrt{N}dm$.

In the thermodynamic limit, $Z_N$ can be evaluated by the saddle point method, so that
\begin{equation}
\mu\left(\Sigma_{v,N}\right)\propto e^{N s_N(\left\langle v\right\rangle,m_0)},
\end{equation}
where $\left\langle v\right\rangle$ is the average potential, and $m_0$ maximizes $s_N(\left\langle v\right\rangle,m)$. Now, it is clear that $\Sigma_{\left\langle v\right\rangle,N}$ to be dumbbell-shaped is a sufficient condition for the $\mathbb{Z}_2$-SB. Furthermore, this condition appears also to be necessary, because if $\Sigma_{\left\langle v\right\rangle,N}$ were not dumbbell-shaped, then the maximum of $s_N(\left\langle v\right\rangle,m)$ would be at $m=0$, so that the $\mathbb{Z}_2$ symmetry would not be broken, against the hypothesis. These consideration, for the sake of mathematical formality, can be condensed in a straightforward theorem.

\medskip
\noindent{\textbf{Theorem 1.}} \emph{Let $(v',v'')$ be an interval of accessible values of the potential density $v$. The $\mathbb{Z}_2$ symmetry is spontaneously broken for suitable values of the temperature $T$ such that $\left\langle v\right\rangle(T)\in (v',v'')$ if, and only if, there exists $N_0\in \mathbb{N}$ such that the $\Sigma_{v,N}$'s are dumbbell-shaped $\forall N>N_0$ and $\forall v\in(v',v'')$.}

\medskip
In the most common case $v'$ is the global minimum of $v$, which is reached at $T=0$.

\medskip
\noindent{\textbf{Definition 2.}} \emph{$\Sigma_{v_c(N),N}$ is \emph{critical} if there exists $v'<v_c(N)$ such that $\Sigma_{v,N}$ is dumbbell-shaped for $v\in [v',v_c(N)]$, and if there exists $v''>v_c(N)$ such that $\Sigma_{v,N}$ is not dumbbell-shaped for $v\in (v_c,v'']$.}

\medskip
\noindent{\textbf{Theorem 2.}} \emph{If there exists $N_0\in\mathbb{N}$ such that $\forall N>N_0$ $\quad\Sigma_{v_c(N),N}$ is critical, then $v_c(N)\to\langle v\rangle_c$ per $N\to\infty$, where $\langle v\rangle_c$ is the critical thermodynamic average potential density.} 

\smallskip
\noindent{Proof.} It is an immediate consequence of the theorem and the definitions given above. $\square$

\medskip
At this point it is worth pointing out two features about the broken phase in the thermodynamic limit.

\smallskip
(i) $s(v,m)=\lim_{N\rightarrow\infty}s_N(v,m)$ is a non-concave $m$-function with two global maxima corresponding to the spontaneous magnetization. Since $s(v,m)$ has to be concave for short-range potentials \cite{g,l}, the generating-mechanism of $\mathbb{Z}_2$-SB seems suitable for long-range potential only.

\smallskip
(ii) $s(v,m)$ does not maintain the non-concavity at finite $N$ of the $s_N(v,m)$'s, as a consequence is a non-strictly concave $m$-function. This is the typical picture of a short-range system, even though a long-range one cannot be "a priori" excluded.

\smallskip
We wonder what a general condition given on the potential globally considered in order to satisfy Theorem 1 may be. As it was already pointed out in Ref. \cite{b2}, a double-well potential may be the most general answer, e.g., the $\phi^4$ model considered in Sec. \ref{phi4}. The global minima of the potential have to be located on a line orthogonal to the $\Sigma_{m,N}$'s defined in (\ref{pim}). This scenario is represented in Fig. \ref{fig_revs_s}. Further, following the example of the hypercubic model, an alternative way is to define the potential foil by foil by shaping the single $\Sigma_{v,N}$ as requested by Theorem 1, even without the presence of the two global minima, but by this way the smoothness cannot be guarantee.

\subsection{On a theorem on a sufficient condition for $\mathbb{Z}_2$-SBPT}

In Ref. \cite{bc} a straightforward theorem (Theorem 1 in the paper) on a sufficient topological condition for $\mathbb{Z}_2$-SBPT was proven. The hypotheses are given in terms of topological properties of the $\Sigma_{v,N}$'s of an $N$ degrees of freedom Hamiltonian system with a $\mathbb{Z}_2$ symmetry. Simplifying a little bit the scenario, the statement is as follows.

\smallskip
\emph{Let $v'<v''$ be two values of the potential density such that $\Sigma_{v,N}=\Sigma_{v,N}^A\cup\Sigma_{v,N}^B$ $\forall v\in (v',v'')$, $\Sigma_{v,N}^A\cap\Sigma_{v,N}^B=\emptyset$, $\Sigma_{v,N}^A\cap\Sigma_{0,N}=\emptyset$ and $\Sigma_{v,N}^B\cap\Sigma_{0,N}=\emptyset$, where $\Sigma_{0,N}$ is the hyperplane at constant magnetization defined in (\ref{pim}), $\Sigma_{v,N}^A\sim\Sigma_{v,N}^B\sim\mathbb{S}^N$, where "$\sim$" stands for "is homeomorphic to". Then, in the thermodynamic limit the $\mathbb{Z}_2$ symmetry is spontaneously broken for the values of temperature $T\in (T',T'')$ such that $v'=\left\langle v\right\rangle(T')$ and $v''=\left\langle v\right\rangle(T'')$.}

\smallskip
In order to show the theorem, it was made the assumption that if the RP is confined in one of the two connected components $\Sigma_{v,N}^A$ and $\Sigma_{v,N}^B$ of $\Sigma_{v,N}$, then the spontaneous magnetization can be calculated by the ensemble average performed only on the connected component where the RP is located. In the light of what discussed here, this is trivially not always true. Nevertheless, the theorem survives as a limiting case of Theorem 1 in Sec. \ref{strangled}. Indeed, since $\Sigma_{v,N}^A\cap\Sigma_{0,N}=\emptyset$ and $\Sigma_{v,N}^B\cap\Sigma_{0,N}=\emptyset$, then $\Sigma_{v,N}$ is trivially dumbbell-shaped and the $\mathbb{Z}_2$ symmetry is broken because of the aforementioned theorem.

\section{Application to a modified version of the hypercubic model}
\label{bs}

\begin{figure}
	\begin{center}
		\includegraphics[width=0.235\textwidth]{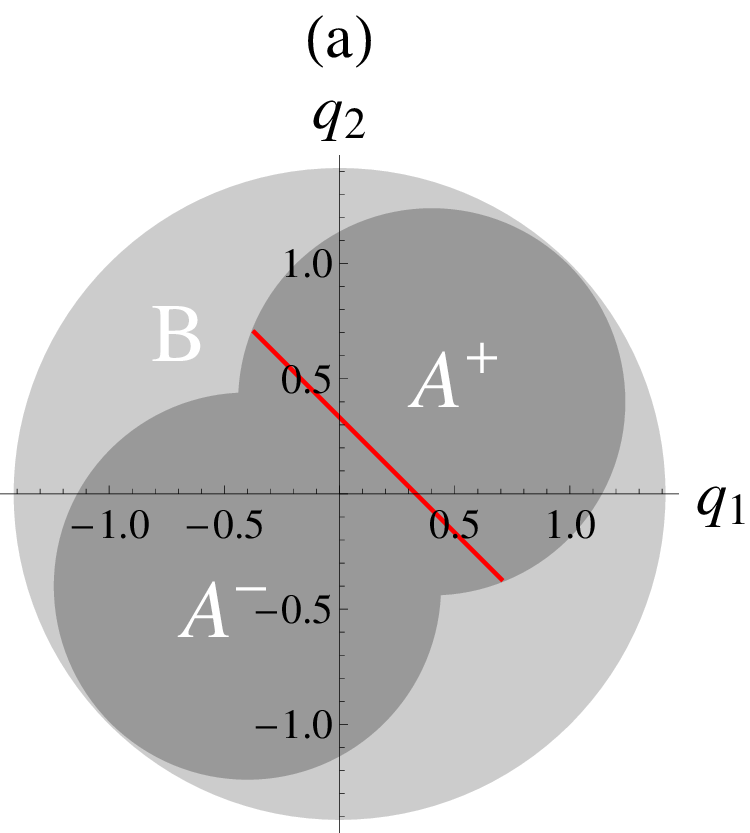}
		\includegraphics[width=0.235\textwidth]{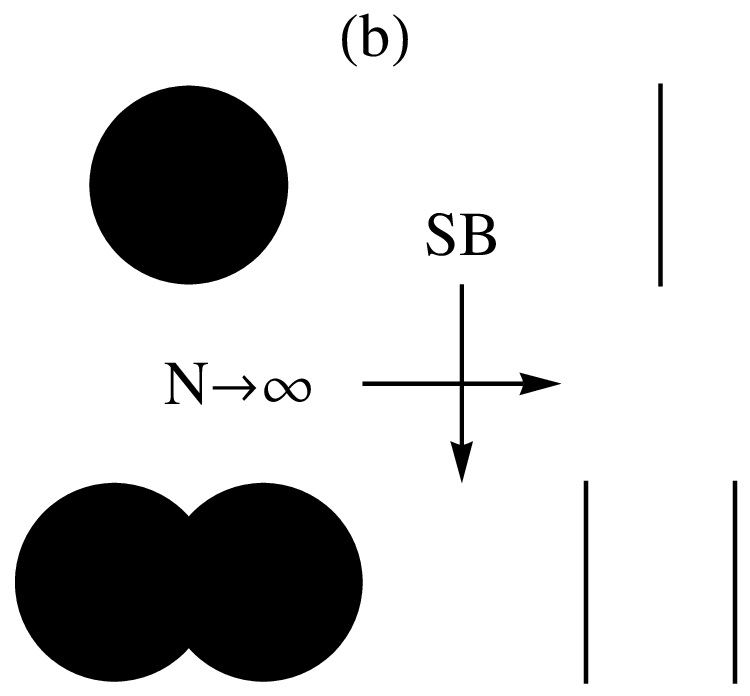}
		\caption{Model of Sec. \ref{bs} with $v_c=1$ and $m_0=0.4$. (a) Sketch of the $\Sigma_{v,2}$'s. $\Sigma_{-v_c,2}=A^+\cup A^-$ and $\Sigma_{0,2}=B\backslash\left(A^+\cup A^-\right)$ where $B$ is the whole $2$-ball. The segment (red) is $A^+\cup A^-\cap\Sigma_{m,2}$. (b) Pictorial representation of the concept of topological limit introduced in Sec. \ref{limtop} of the $\Sigma_{v,N}$'s. Only in this limit we can associate a topological change with the $\mathbb{Z}_2$-SB.}
		\label{ballesecanti}
	\end{center}
\end{figure}

Unfortunately, the calculation of $A^+\cup A^-\cap\Sigma_{m,N}$ of the hypercubic model (\ref{Vhcm}) is not analytically feasible. 
Therefore, we replace the $N$-cubes by $N$-balls, as sketched in Fig. \ref{ballesecanti}. The radius of $B$ is assumed to be $\sqrt{N}$ to give rise to a magnetization $m\in [-1,1]$, while the radius of $A^+$, $A^-$ is assumed to be  $(1-m_0)\sqrt{N}$, where $m_0$ is the value of the spontaneous magnetization. $0<m_0<1$ is assumed. As for the hypercubic model, the potential takes only the values $-v_c,0$. 

\begin{figure}
	\begin{center}
		\includegraphics[width=0.235\textwidth]{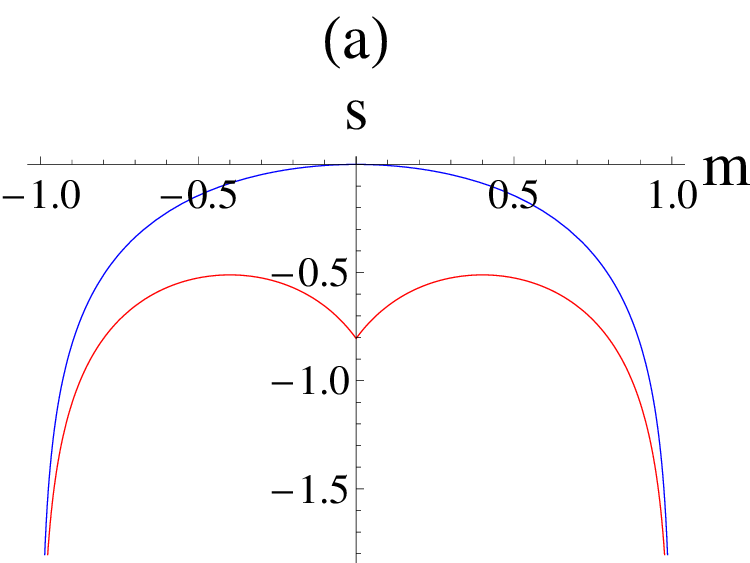}
		\includegraphics[width=0.235\textwidth]{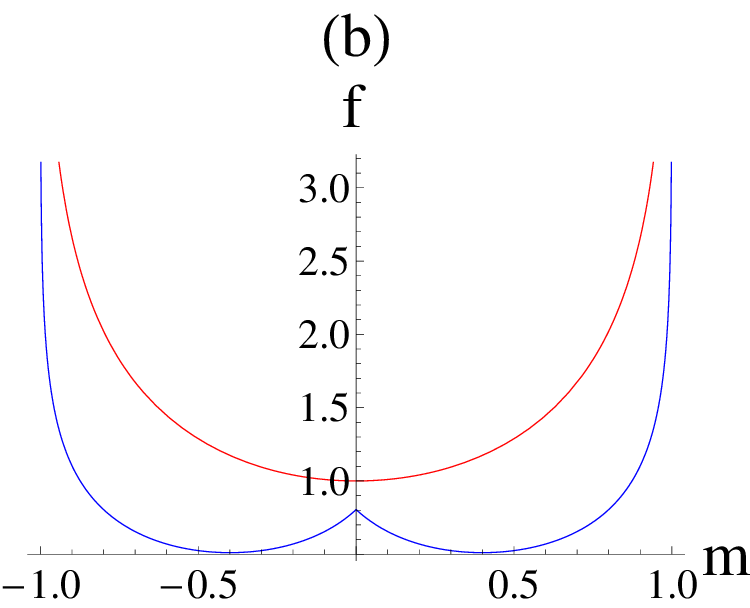}
		\caption{Model of Sec. \ref{bs} with $v_c=1$ and $m_0=0.4$. (a) Microcanonical entropy $s(v,m)$ at fixed $v=0$ (smooth blue line), and at $v=-v_c$ (red line). (b) As panel (a) for the free energy $f(m,T)$ at fixed $T<T_c$ (blue line) and at $T>T_c$ (smooth red line).}
		\label{bs_smfm}
	\end{center}
\end{figure}

To calculate the density of states as an $m$-function we start from $v=v_c$, thus 
\begin{equation}
\omega_N(-v_c,m)=\mu\left(\Sigma_{-v_c,N}\cap\Sigma_{m,N}\right)=\mu\left(A^+\cup A^-\cap\Sigma_{m,N}\right).
\end{equation}
$A^+\cup A^-\cap\Sigma_{m,N}$ is an $(N-1)$-ball whose radius is given by 
\begin{equation}
r(m)=\sqrt{N}\left((1-m_0)^2-(|m|-m_0)^2\right)^{\frac{1}{2}}.
\end{equation}
Thus, the density of states at fixed $m$ is the volume of the $(N-1)$-ball of radius $r(m)$
\begin{equation}
\omega_N(-v_c,m)=\frac{\pi^{\frac{N}{2}}N^{\frac{N-1}{2}}}{\Gamma\left(\frac{N}{2}+1\right)}\left((1-m_0)^2-(|m|-m_0)^2\right)^{\frac{N-1}{2}}.
\end{equation}
Finally, the entropy in the thermodynamic limit is given by 
\begin{eqnarray}
s(-v_c,m)&=&\lim_{N\to\infty}\frac{1}{N}\ln \omega_N(-v_c,m)\nonumber
\\
&=&\frac{1}{2}\ln\left((1-m_0)^2-(|m|-m_0)^2\right).
\label{svcmbs}
\end{eqnarray}
See Fig. \ref{bs_smfm} for a plot. $s(-v_c,m)$ shows a local minimum at $m=0$ and two global maxima at $m=\pm m_0$ corresponding to the spontaneous magnetization. The fact that $A^+\cup A^-$ is dumbbell-shaped has reflected in a non-concave entropy entailing the spontaneous $\mathbb{Z}_2$-SB.

Now, consider the case $v=0$:
\begin{eqnarray}
\omega_N(0,m)&=&\mu\left(\Sigma_{0,N}\cap\Sigma_{m,N}\right)\nonumber
\\
&=&\mu\left(\left(B\backslash\left(A^+\cup A^-\right)\right)\cap\Sigma_{m,N}\right)\nonumber
\\
&=&\mu\left(\left(B\cap\Sigma_{m,N}\right)\backslash\left(A^+\cup A^-\cap\Sigma_{m,N}\right)\right).
\end{eqnarray}
Since the radius of $A^+\cup A^-\cap\Sigma_{m,N}$ is less than that of $B\cap\Sigma_{m,N}$, the contribution of the former is vanishing for $N\rightarrow\infty$ with respect to the latter, so that
\begin{eqnarray}
\omega_N(0,m)&\backsimeq &\mu\left(B\cap\Sigma_{m,N}\right)\nonumber
\\
&=&\frac{\pi^{\frac{N}{2}}N^{\frac{N-1}{2}}}{\Gamma\left(\frac{N}{2}+1\right)}\left(1-|m|\right)^{N-1},
\end{eqnarray}
and finally, the entropy is
\begin{eqnarray}
s(0,m)=\lim_{N\to\infty}\frac{1}{N}\ln \omega_N(0,m)=\ln\left(1-|m|\right).
\label{s0mbs}
\end{eqnarray}

The importance of this model is in showing, in a discrete and elementary way, the generating-mechanism of a $\mathbb{Z}_2$-SBPT based on the framework of dumbbell-shaped $\Sigma_v$'s. In the general case this occurs in a continuous way giving rise to a continuous $\mathbb{Z}_2$-SBPT, as we show by the models of next sections.

\subsection{Topological limit of $\Sigma_{v,N}$ for large $N$ and asymptotic diffeomorphicity}
\label{limtop}

In Refs. \cite{pettini,acprz,b,bc,ccp1,ckn,gss,gm,k,dgppdfv,pgfcp,rs} a big effort was made to discover any possible link between the topological changes, if any, of the $\Sigma_{v,N}$'s occurring at the critical thermodynamic average potential of the PT and the PT itself. The leading idea, called \emph{topological hypothesis}, is that the PT is generated by some suitable topological changes. The results found here show that a $\mathbb{Z}_2$-SBPT is related to the criticality of the $\Sigma_{v,N}$'s according to Definition 2 in Sec. \ref{strangled}, which is largely independent of the their topology. 

However, in the framework of dumbbell-shaped $\Sigma_{v,N}$'s we can restore, in some extent, the original idea of the topological hypothesis by introducing the concept of "topological limit" of a $\Sigma_{v,N}$ as $N\to\infty$. For precision, we do not should speak about the limit for $N\to\infty$, but the limit for large $N$ because in the former case no $\Sigma_{v,N}$ longer exists. We specify that the definition given here of topological limit does not claim rigor from the mathematical point of view, but only serves to illustrate a concept.

We give the definition of topological limit by the example of the model of Sec. \ref{bs}. In Sec. \ref{strangled} we have showed that only the subset of $\Sigma_{v,N}$ where the microcanonical entropy takes the global maximum gives the leading contribution to the probability of finding the RP in configuration space, so that all the remainder parts of $\Sigma_{v,N}$ can be disregarded as $N\to\infty$. Distinguish two cases:

\smallskip
(i) $v=-v_c$. $\Sigma_{-v_c,N}$ is the non-disjointed union of two $N$-balls, so that the $\Sigma_{-v_c,N}$'s are topologically equivalent to an $N$-ball alone. From Eq. (\ref{svcmbs}) we note that the entropy has two global maxima at $m=\pm m_0$. Thus, we assume the disjointed union of the two $(N-1)$-balls $\Sigma_{-v_c,N}\cap\Sigma_{\pm m_0,N}$ to be the topological limit of $\Sigma_{-v_c,N}$.

\smallskip
(ii) $v=0$. $\Sigma_{0,N}$ is an $N$-ball centered in the origin of configuration space. From Eq. (\ref{s0mbs}) we note that the global maximum of the entropy is at $m=0$. Thus, the $(N-1)$-ball $\Sigma_{-v_c,N}\cap\Sigma_{0,N}$ is assumed to be the topological limit of the $\Sigma_{0,N}$'s.

\smallskip
The fundamental thing is that the topological limit of the sequence of the $\Sigma_{-v_c,N}$'s is not equivalent to their topology. Things go as if a topological change occurs in the thermodynamic limit, but this does not occur in the $\Sigma_{0,N}$'s. From the viewpoint of the $\mathbb{Z}_2$-SBPT, a limiting topological change is exactly located in correspondence of the thermodynamic critical potential. A pictorial representation is given in panel (b) of Fig. \ref{ballesecanti}.

In Refs. \cite{gfp,gfp1} the concept of "asymptotic diffeomorphicity" was put forward. To have an intuitive idea of how asymptotic diffeomorphicity works, we refer to Fig. 1 in Ref. \cite{gfp} and Fig. 4 in Ref. \cite{gfp1}. In our opinion, this picture is very similar to what we have expressed in this section by the concept of topological limit. 

To give an example, consider the 2d $\phi^4$ model with neighbor-nearest interaction studied in Refs. \cite{gfp,gfp1}. The $\Sigma_{v,N}$'s are diffeomorphic to an $N$-sphere for $v>0$. For a suitable choice of the free parameters of the model, $\langle v\rangle_c>0$ holds. This means that no topology change occurs while crossing $\langle v\rangle_c$. From the viewpoint of asymptotic diffeomorphicity, instead, a topology change occurs just at $v=\langle v\rangle_c$. Indeed, for $0<v<\langle v\rangle_c$ the $\Sigma_{v,N}$'s are not asymptotically diffeomorphic to an $N$-sphere, but to two $N$-spheres, while for $v>\langle v\rangle_c$ the $\Sigma_{v,N}$'s are asymptotically diffeomorphic to an $N$-sphere, just as for any $N$. This scenario is quite similar to that of dumbbell-shaped $\Sigma_{v,N}$'s. We ask weather a logical equivalence between the two concepts, or at least a less stringent logical relationship, may hold. The presence of the neck in the $\Sigma_{v,N}$'s for $0<v<\langle v\rangle_c$ may be the key element that breaks the asymptotic diffeomorphicity. Investigating this could be an interesting line of future research.

\section{Application to a model with a classical $\mathbb{Z}_2$-SBPT}
\label{revolution}

In Ref. \cite{b4} a model to which the dumbbell-shaped $\Sigma_v$'s method can be applied was introduced. The potential does not describe any physical system, nevertheless, has all the characteristics of a physical system with a $\mathbb{Z}_2$-SBPT with classical critical exponents. Since the model has also an $O(N-1)$ symmetry around the line passing through the origin and perpendicular to the $\Sigma_{m,N}$'s, it has been called \emph{revolution model}. The advantage of this model is that $\omega_N(v,m)$ can be analytically calculated.  

Let $(q_1,\cdots,q_{N})$ be the standard coordinate system of $\mathbb{R}^N$. The starting point to define the potential is setting the new coordinate system
\begin{equation}
(m,\widetilde{q}_1,\cdots,\widetilde{q}_{N-1}),
\label{cs}
\end{equation}
where the direction of the first coordinate $m$ is the line orthogonal to the $\Sigma_{m,N}$'s and passing through the origin, and $(\widetilde{q}_1,\cdots,\widetilde{q}_{N-1})$ is an orthonormal coordinate set orthogonal to $m$. 

The potential is defined as
\begin{equation}
V=N(-Jm^2+m^4)+e^{2m^2}\sum_{i=1}^{N-1}\tilde{q}_i^2.
\label{Vrevs}
\end{equation}
The constant $J\ge 0$ plays the role of a coupling constant. The canonical thermodynamic can be solved analytically, and the critical temperature is $T_c=J$. For our purposes, we limit to report the spontaneous magnetization and the specific average potential, respectively,
\begin{equation}
\left\langle m\right\rangle=\left\{\begin{array}{ll}
\pm\frac{1}{\sqrt{2}}(J-T)^{\frac{1}{2}}\quad &\hbox{if} \quad T\le T_c
\\
0\quad &\hbox{if} \quad T\ge T_c
\end{array} \right.,
\label{mT_revs}
\end{equation}
\begin{equation}
\left\langle v\right\rangle=\left\{\begin{array}{ll}
\frac{T}{2}-\frac{1}{4}(J-T^2)\quad &\hbox{if} \quad T\le T_c
\\
\frac{J}{2}\quad &\hbox{if} \quad T\ge T_c
\end{array} \right..
\label{vT_revs}
\end{equation}

\subsection{Dumbbell-shaped $\Sigma_v$'s}

The potential (\ref{Vrevs}) has two global minima of value $-NJ/4$ whose coordinates are $\left(\pm\sqrt{N/2},0,\cdots,0\right)$, and has a saddle point of value $0$ at $(0,\cdots,0)$. 

The topology of the $\Sigma_{v,N}$'s is as follows ("$\sim$" stands for "is homeomorphic to")
\begin{equation}
\Sigma_{v,N}\sim\left\{\begin{array}{ll}
\mathbb{S}^{N-1}\quad &\hbox{if} \quad v>0
\\
\hbox{critical} &\hbox{if} \quad v=0
\\
\mathbb{S}^{N-1}\cup \mathbb{S}^{N-1} &\hbox{if} \quad 0>v\geq-\frac{J}{4}
\\
\emptyset &\hbox{if} \quad v<-\frac{J}{4}
\end{array}\right..
\end{equation}
There exists only a topological change at $v=0$. The potential satisfy the hypotheses of Theorem $1$ in Ref. \cite{bc} for $v\in [-J/4,0)$, so that the $\mathbb{Z}_2$-SB is guaranteed for $T\in [0,T')$ by topological reasons, where $T'=\left\langle v\right\rangle^{-1}(0)=-1+\sqrt{1+J}$ is obtained by inverting the relation (\ref{vT_revs}). Indeed, the $\Sigma_{v,N}$'s are made up by two connected components which are one the image of the other under the $\mathbb{Z}_2$ symmetry.

The critical average potential $\langle v\rangle_c=J/2$ is located above $0$, to which  the unique critical $v$-level set $\Sigma_{0,N}$ corresponds. This is due to the dumbbell-shaped $\Sigma_{v,N}$'s in the interval $[0,v_c)$ which, according to Theorem 1 in Sec. \ref{strangled}, imply the $\mathbb{Z}_2$-SB. 

The simplicity of this model, in particular the presence of the $O(N-1)$ symmetry, allows us to identify the dumbbell-shaped $\Sigma_{v,N}$'s by the analytic calculation of the density of states $\omega_N(v,m)=\mu\left(\Sigma_{v,N}\cap\Sigma_{m,N}\right)$. Indeed, $\Sigma_{v,N}\cap\Sigma_{m,N}$ is an $(N-1)$-sphere defined by the following equation
\begin{equation}
Nv=N(-Jm^2+m^4)+e^{2m^2}\sum_{i=1}^{N-1}\tilde{q}_i^2,
\end{equation}
whose radius $R$ is given by
\begin{equation}
R^2=\sum_{i=1}^{N-1}\tilde{q}_i^2=Ne^{-2m^2}(v+Jm^2-m^4),
\end{equation}
and whose volume is given by
\begin{equation}
vol\left(\Sigma_{v,N}\cap\Sigma_{m,N}\right)=\frac{2\pi^{\frac{N-1}{2}}}{\Gamma\left(\frac{N-1}{2}\right)}R^{N-2}.
\end{equation}
Here, we limit to report only the final result of the calculation of the microcanonical entropy which, in the limit $N\to\infty$, result to be
\begin{eqnarray}
& &s(v,m)=\lim_{N\rightarrow\infty}\ln\omega_N(v,m)^{\frac{1}{N}}=\nonumber
\\
& &=-m^2+\frac{1}{2}\ln(v-m^4+Jm^2)+\frac{1}{2}\ln(2\pi e).
\label{revs_a}
\end{eqnarray}
See Fig. \ref{fig_revs_s} for some plots.
\begin{figure}
	\begin{center}
		\includegraphics[width=0.235\textwidth]{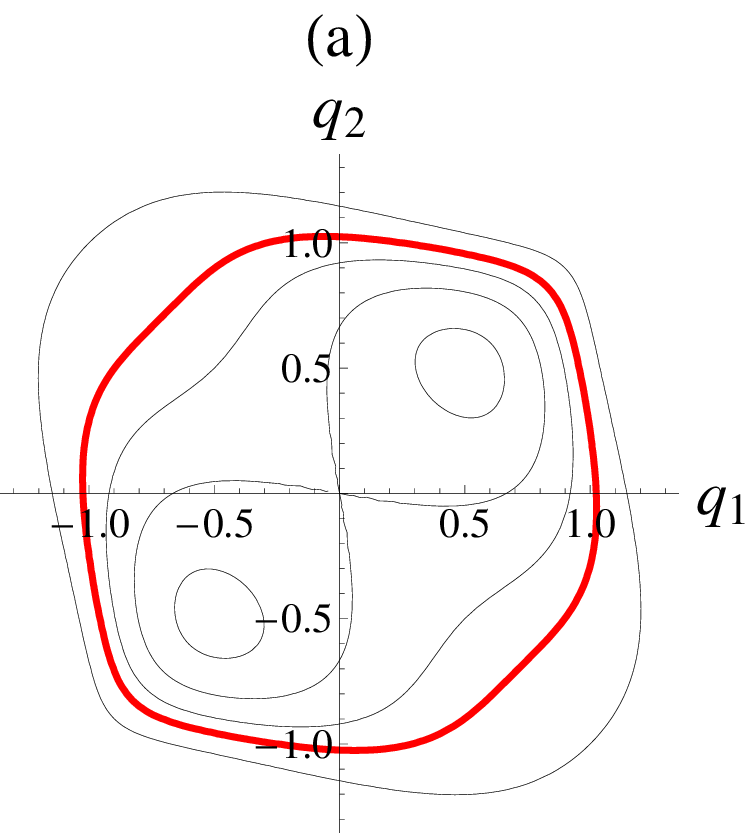}
		\includegraphics[width=0.235\textwidth]{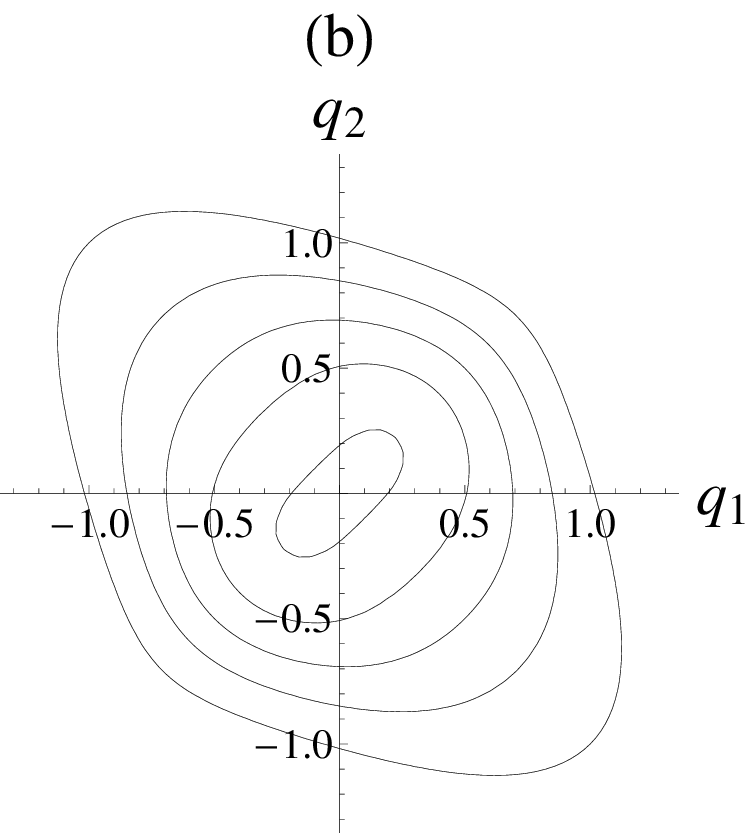}
		\caption{(a) Some $\Sigma_{v,2}$'s of the revolution model (\ref{Vrevs}) for $J=1$ and $v=-0.2, ~0, ~0.25 ,~0.5,~ 1$ starting from the innermost one, respectively. $\Sigma_{0.5,2}$ (thick red) is the boundary between the dumbbell-shaped $\Sigma_{v,2}$'s for $v\in [-0.25,0.5]$ and those which are not dumbbell-shaped for $v\geq 0.5$. (b) The same of panel (a) for $J=0$ and $v=0.01, ~0.1, ~0.25 ,~0.5,~ 1$.}
		\label{fig_revs_sigmav}
	\end{center}
\end{figure}

According to Definition 1 in Sec. \ref{strangled}, a $\Sigma_{v,N}$ is dumbbell-shaped if the restriction of $s(v,m)$ on $\Sigma_{v,N}$ as an $m$-function does not take the global maximum at $m=0$. For $v\in[-J/4,0)$ the $\Sigma_{v,N}$'s are dumbbell-shaped because they are the union of two connected components (see Fig. \ref{fig_revs_sigmav}). The solution with respect to $v$ of the following equation
\begin{equation}
\frac{\partial s(v,m)}{\partial m}=0
\end{equation}
gives the spontaneous magnetization as a $v$-function 
\begin{equation}
m(v)=\begin{cases}
\pm\left(1-\left(v+\frac{1}{2}\right)^\frac{1}{2}\right)^\frac{1}{2} & \text{ if }\quad -\frac{1}{4}\le v\le \frac{1}{2}
\\
0 & \text{ if } \quad v\ge\frac{1}{2}
\end{cases}.
\end{equation}
By inserting Eq. (\ref{vT_revs}) in the last one we get Eq. (\ref{mT_revs}).

Consider $v\geq 0$. To discover weather a $\Sigma_{v,N}$ is dumbbell-shaped is sufficient to set to zero the second partial derivative of $s(v,m)$ with respect to $m$ at $m=0$
\begin{equation}
\left.\frac{\partial^2 s(v,m)}{\partial m^2}\right|_{m=0}=2 v-J=0.
\end{equation}
Therefore, $v=J/2$ is the boundary between the dumbbell-shaped $\Sigma_{v,N}$'s from those which are not dumbbell-shaped. In particular, the $\Sigma_{v,N}$'s are dumbbell-shaped for $v<J/2$ (see Fig. \ref{fig_revs_sigmav}). $\Sigma_{J/2,N}$ is critical according to Definition 2 in Sec. \ref{strangled}.
\begin{figure}
	\begin{center}
		\includegraphics[width=0.235\textwidth]{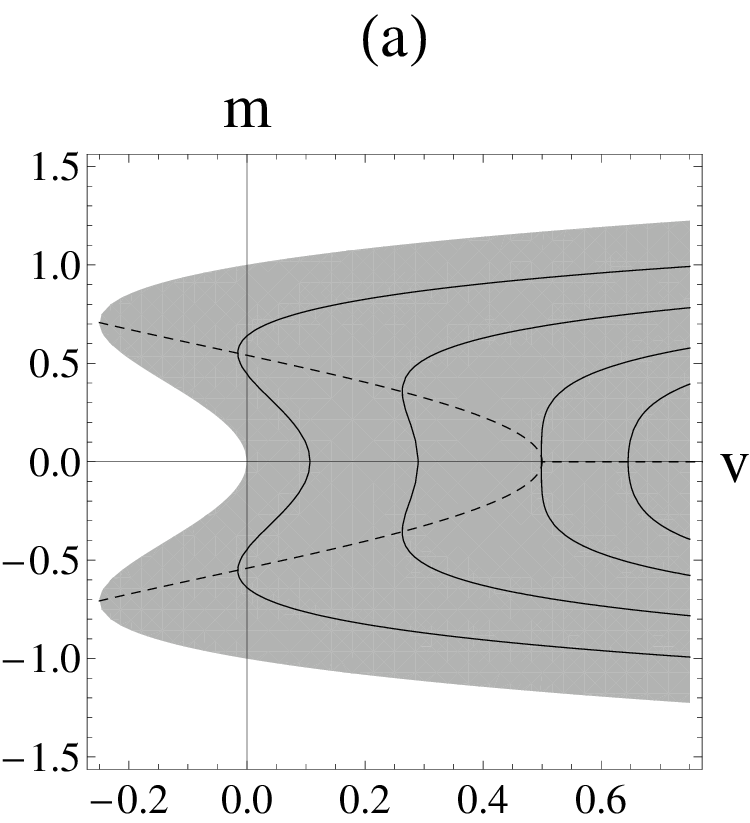}
		\includegraphics[width=0.235\textwidth]{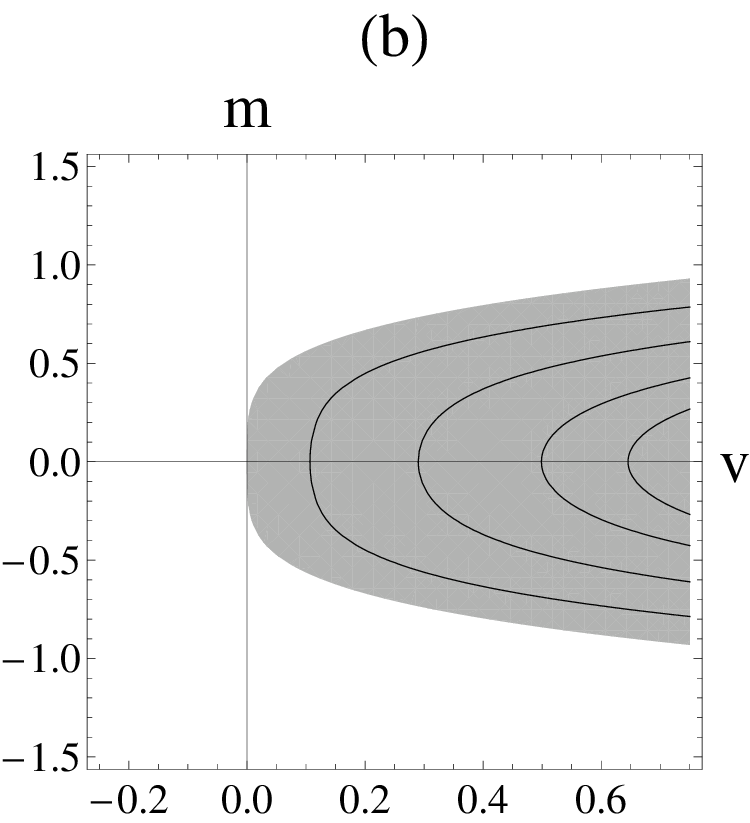}
		\caption{(a) Revolution model (\ref{Vrevs}) for $J=1$. Contour plot of the microcanonical entropy $s(v,m)$ (\ref{revs_a}), the dark region surrounded by the curve of equation $v=-Jm^2+m^4$ is the domain of $s(v,m)$. $v=\langle v\rangle_c=0.5$ is the boundary between the dumbbell-shaped $\Sigma_{v,N}$'s and the non-dumbbell-shaped ones. (b) The same of panel (a) for $J=0$.}
		\label{fig_revs_s}
	\end{center}
\end{figure}
From a thermodynamic viewpoint, the critical average potential is just $\langle v\rangle_c=\left\langle v\right\rangle(T_c)=J/2$. 

Summarizing, the thermodynamic picture of the $\mathbb{Z}_2$-SBPT is in perfect agreement with the geometric picture of the dumbbell-shaped $\Sigma_{v,N}$'s introduced in Sec. \ref{strangled}. It is a remarkable fact that we can get the same result in two independent ways.

\subsection{The case at finite N}
\label{revsN}

The formula of the microcanonical entropy for finite $N$ is the following
\begin{equation}
\begin{split}
s_N(v,m)=-\frac{N-5}{N}m^2+
\frac{N-5}{2N}\ln N+\frac{N-1}{2N}\ln\pi+\\
\frac{N-3}{2N}\ln\left(v-m^4+Jm^2\right)-\frac{1}{N}\ln\Gamma\left(\frac{N-1}{2}\right).
\end{split}
\label{revs_sN}
\end{equation}
There are no substantial differences in the shape of the graph compared to the case $N=\infty$ discussed in previous section. The definition of $\mathbb{Z}_2$-SBPT given in Sec. \ref{strangled} holds also for finite $N$. Therefore, the model (\ref{Vrevs}) undergoes the spontaneous symmetry breaking of its $\mathbb{Z}_2$ symmetry for every $N>5$. The smaller $N$'s must be excluded because Eq. (\ref{revs_sN}) makes no sense. The critical potential 
\begin{equation}
v_c(N)=\frac{J (N-3)}{2 (N-5)}.
\end{equation}
is an $N$-function tending to $\langle v_c\rangle=1/2$ for $N\to\infty$, as expected from Theorem 2 in Sec. \ref{strangled}.

We do not enter into the discussion of what the temperature for finite $N$ may be, in particular the critical temperature, because this is not the right place to deal with this problem. We merely observe that we do not see any substantial difference with respect to the thermodynamic limit except in the fact that the fluctuations of the physical quantities are non-vanishing.
\begin{figure}
	\begin{center}
		\includegraphics[width=0.35\textwidth]{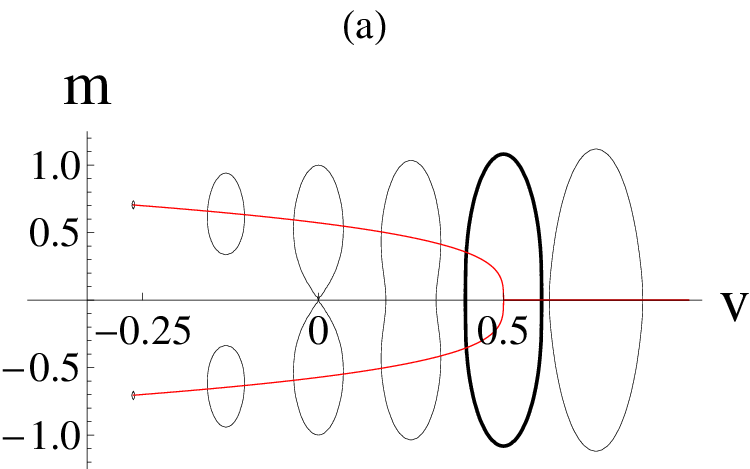}
		\includegraphics[width=0.35\textwidth]{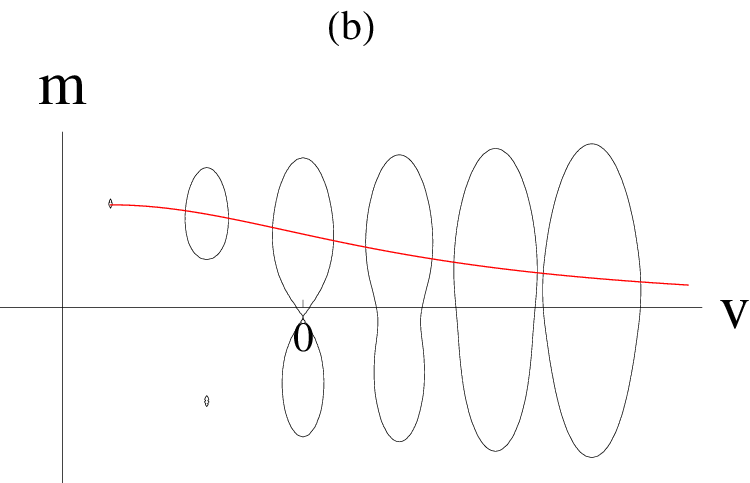}
		\caption{(a) Revolution model (\ref{Vrevs}) for $J=1$. Pictorial representation of the relation between the spontaneous magnetization (red lines) and the shape of the $\Sigma_{v,N}$'s. The thick one is critical according to Definition 2 in Sec. \ref{strangled}. (b) The same of panel (a) for a non-zero external magnetic field which breaks the $\mathbb{Z}_2$ symmetry.}
		\label{magsigmav}
	\end{center}
\end{figure}

\section{Application to a physical model: the mean-field $\phi^4$ model}
\label{phi4}

We recall the potential of the mean-field $\phi^4$ model
\begin{equation}
V=\sum^{N}_{i=1}\left(-\frac{\phi_i^2}{2}+\frac{\phi_i^4}{4}\right)-\frac{J}{2N}\left(\sum^{N}_{i=1}\phi_i\right)^2.
\label{Vphi4}
\end{equation}
The model is known to undergo a $\mathbb{Z}_2$-SBPT with classical critical exponents.

In Ref. \cite{hk} the authors were able to calculate the thermodynamic limit of the microcanonical entropy $s(v,m)$ by the large deviations theory. The domain of $s(v,m)$ is a non-convex subset of the plane $(v,m)$, and $s(v,m)$ is a non-concave function, coherently with the long-range interaction of the model. The critical average potential $\langle v\rangle_c$ of the $\mathbb{Z}_2$-SBPT is located in such a way to divide the concave sections $s(v,m)$ at fixed $v$ at $v\ge \langle v\rangle_c$ from the non-concave ones at $v<\langle v\rangle_c$. The graphic of $s(v,m)$ in Ref. \cite{hk} is qualitatively identical to that of the revolution model (\ref{Vrevs}) plotted in Fig. \ref{fig_revs_s}.

In Refs. \cite{aarz,b0,gss} the topology of the $\Sigma_{v,N}$'s was exhaustively studied by Morse theory \cite{pettini}. The following three cases have delineated:

\smallskip
(i) $v\in [v_{min},v_t)$, where $v_{min}=-(1+J)^2/4$ is the global minimum of the potential. $v_t$ depends on the coupling constant $J$, and $v_t<-1/4$. The $\Sigma_{v,N}$'s are homeomorphic to the union of two disjoint $N$-spheres. The critical potential of the $\mathbb{Z}_2$-SBPT may be less than $0$, but $v_c>v_t$ holds for every $J$. 

\smallskip
(ii) $v\in [v_t,0]$. There is a huge amount of critical points growing as $e^N$. We can say that the whole interval $[v_t,0]$ plays the role of a critical $v$-level set which discriminates between the $\Sigma_{v,N}$'s homeomorphic to two disjointed $N$-spheres from the ones homeomorphic to an $N$-sphere alone. In the following section we see how it is possible to reduce this critical interval to a single critical $\Sigma_{v,N}$ containing a single critical point. Furthermore, $v_t\rightarrow-1/4^-$ as $J\rightarrow\infty$,.

\smallskip
(iii) $v\in (0,+\infty)$. The $\Sigma_{v,N}$'s are homeomorphic to an $N$-sphere. 

\smallskip
The interpretation of this scenario leads to the same conclusion depicted in Sec. \ref{revolution} for the revolution model (\ref{Vrevs}). Let us consider the three cases in detail. 

In the case (i) the hypotheses of Theorem 1 in Ref. \cite{bc} are satisfied, thus the topology of the $\Sigma_{v,N}$'s implies the $\mathbb{Z}_2$-SB. This is in accordance with $\langle v\rangle_c>v_t$ for every $J$, because the magnetization cannot vanish below $v_t$. As showed in Sec. \ref{strangled}, since Theorem 1 in Ref. \cite{bc} is a special case of Theorem 1 in Sec. \ref{strangled}, also the hypotheses of the latter are satisfied. 

In the case (ii) the hypotheses of Theorem 1 in Ref. \cite{bc} are not satisfied, so that only Theorem 1 in Sec. \ref{strangled} can implies the $\mathbb{Z}_2$-SB, because the $\Sigma_{v,N}$'s may be dumbbell-shaped below $\langle v\rangle_c$ and not any more above $\langle v\rangle_c$ (if $\langle v\rangle_c<0$) independently on their intricate topology.

Finally, the same of the case (ii) holds for the case (iii), with the non-significant difference that the $\Sigma_{v,N}$'s are all diffeomorphic to an $N$-sphere.

From Definition 2 and Theorem 2 in Sec. \ref{strangled}, we aspect that at fixed $N$ there exists $v_c(N)$ such that $\Sigma_{v_c(N),N}$ is critical and  $v_c(N)\rightarrow \langle v\rangle_c$ for $N\rightarrow\infty$. Further analytic and numerical studies may check this conjecture.

\subsection{A simplified version of the mean-field $\phi^4$ model}
\label{pphi4}

In Ref. \cite{b_5} a simplified version of the $\phi^4$ model was introduced and studied in the mean-field version. The simplification is nothing more than the elimination of the quadratic term in the local potential. The new potential is therefore the following
\begin{equation}
V=\sum^{N}_{i=1}\frac{\phi_i^4}{4}-\frac{J}{2N}\left(\sum^{N}_{i=1}\phi_i\right)^2.
\label{Vpphi4}
\end{equation}
It was shown that the quadratic term has no role in generating the $\mathbb{Z}_2$-SBPT, which is identical to that of the traditional model apart quantitative differences. On the other hand, the quadratic term is a cause of great complication in the topological structure of the $\Sigma_{v,N}$'s, which has been described in previous section. Thanks to this simplification, the potential landscape undergoes a topological trivialization. Indeed, only three critical points survive against the immense multitude of the model with non-vanishing quadratic term. We speak of "topological trivialization" because to have a double-well potential, at least two global minima with index $0$ and a central saddle point with index $1$ are needed.
\begin{figure}
	\begin{center}
		\includegraphics[width=0.235\textwidth]{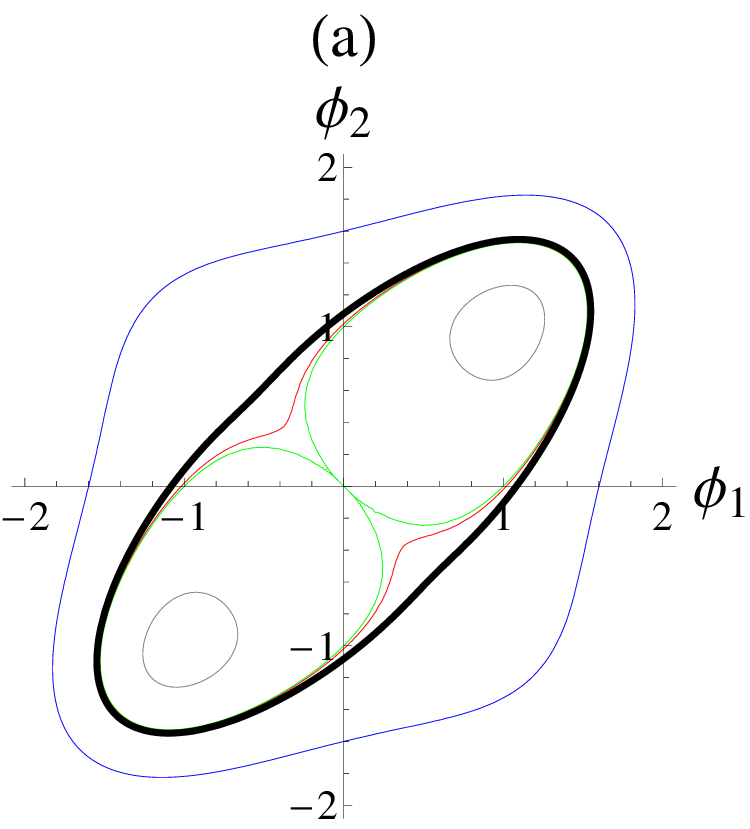}
		\includegraphics[width=0.235\textwidth]{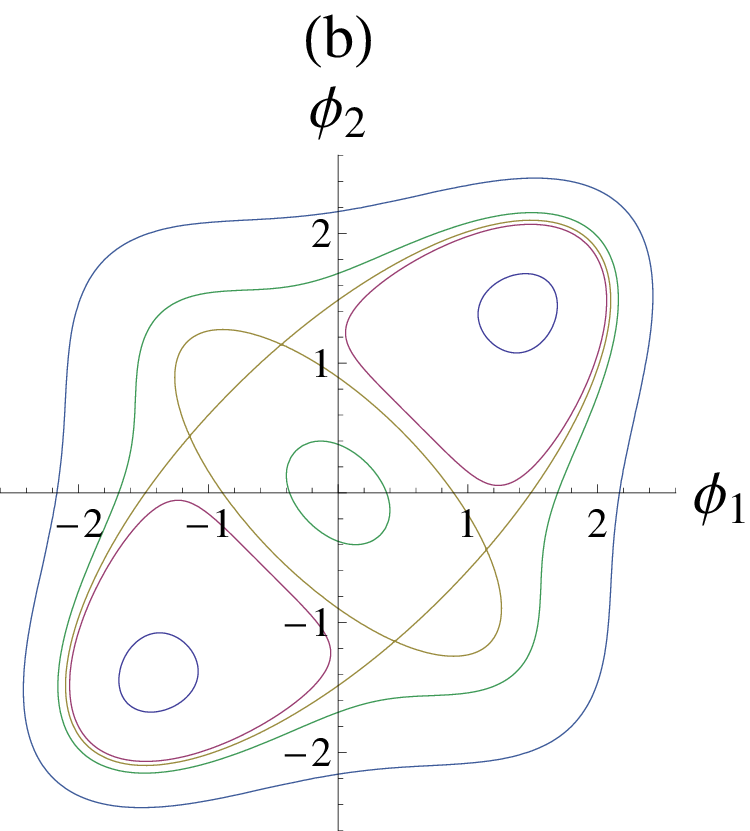}
		\caption{(a) Some $\Sigma_{v,2}$'s of the mean-field  simplified $\phi^4$ model (\ref{Vpphi4}) for $J=1$ and $v=-0.4, ~0, ~0.01 ,~0.05,~ 1$ starting from the innermost one, respectively. $\Sigma_{0.05,2}$ (thick) is the boundary between the dumbbell-shaped $\Sigma_{v,N}$'s and those which are not dumbbell-shaped. $0.05$ is only a numerical estimate because it is not possible to evaluate it analytically. (b) The same of panel (a) for the mean-field $\phi^4$ model (\ref{Vphi4}) for  $J=1$ and $v=-1.8, ~-0.6, ~-0.4375 ,~-0.1,~ 2$. The proliferation of the critical points is evident.}
		\label{pphi4phi4}
	\end{center}
\end{figure}
In Fig. \ref{pphi4phi4} the proliferation of the critical points in the mean-field $\phi^4$ model due to the non-vanishing quadratic term of the local potential is already evident at $N=2$.

In the case of the mean-field simplified $\phi^4$ model the connection between the interpretation of $\mathbb{Z}_2$-SBPTs proposed in this paper and a physical model becomes truly evident. We want to emphasize that the potentials of this model and the model (\ref{Vrevs}) have the same topological and geometric structure, indeed they can be transformed one into the other by a mere shape variation. A comparison between panel (a) of Fig. \ref{fig_revs_sigmav} and panel (a) of Fig. \ref{pphi4phi4} clarifies this at $N=2$. The difference is that only in the case of the model (\ref{Vrevs}) the analytical calculation of the microcanonical entropy $s(v,m)$ is feasible. In the case of the mean-field simplified $\phi^4$ model the large deviation theory may be applied as it was made in Ref. \cite{hk} for the traditional model.

\subsection{On the origin of PTs}

In Refs. \cite{pettini,acprz,b,bc,ccp1,ckn,gss,gm,k,rs} a great effort has been put in trying to understand the origin of a PT meant as a loss of analyticity in the thermodynamic functions in the light of the topological hypothesis. Here, our purpose is to make some trivial considerations about that question. 

The free energy $f$ in the $(m,T)$-plane is an analytic function in the thermodynamic limit, but, e.g., this is not the case of the spontaneous magnetization as a $T$-function. By resorting to Dini's theorem (or the implicit function theorem) we know that the graph of the zeroes of the partial derivative with respect to $m$ of $f$, i.e., the spontaneous magnetization, is an analytic function too. More precisely, if $f(m,T)\in C^k$, then also $m(T)\in C^k$ for $k=1,\cdots,\infty$. 

For example, consider the model (\ref{Vrevs}). The singular point in the graph of $m(T)$ arises because it is the union of two analytic branches which cannot be jointed without passing through a non-analytic point. The two analytic branches are the line $m(T)=0$ for $T\geq 0$, and the parabola $m(T)=\pm(T_c-T)^{1/2}$ for $T\leq T_c$, which intersect at $(0,T_c)$. In Ref. \cite{hk} it was shown that a non-analyticity in the entropy $s(v)$ stems from a maximization-process of the entropy $s(v,m)$ with respect to $m$. This fact is strictly correlated to what aforementioned because it holds only if the graph of $s(v,m)$ is non-concave. There is no way to generate such a non-analyticity starting from a strictly concave graph. 

A non-strictly concave graph is a typical presence when a PT is accompanied with $\mathbb{Z}_2$-SB, but we wonder if this picture may be extended when this does not hold. For example, in the hypercubic model (\ref{Vhcm}), the first-order PT is not related to the $\mathbb{Z}_2$-SB, but rather it stems from the fact that the potential is not a continuous function of the coordinates. Indeed, it assumes only some discrete values. To conclude, at this stage we cannot suggest any unified origin for a PT to occur.

\section{Concluding remarks}

In this paper we have shown a necessary and sufficient condition for a $\mathbb{Z}_2$-SBPT to occur in Hamiltonian systems. This condition is based on a particular shape of the $\Sigma_{v,N}$'s that we have called \emph{dumbbell-shaped}, i.e., made by two lobes connected by a narrow neck. A dumbbell-shaped $\Sigma_{v,N}$ entails the spontaneous ergodicity breaking, whence the $\mathbb{Z}_2$-SBPT. Since in the limit of large $N$ the statistical measure shrinks around $\Sigma_{\langle v\rangle (T),N}$, so that the last one is the unique equipotential surface accessible to the RP with good approximation, we can calculate the spontaneous magnetization by the ensemble average on the $\Sigma_{\langle v\rangle (T),N}$ provided that the last one is ergodic. 

But we have shown that if $\Sigma_{\langle v\rangle (T),N}$ is dumbbell-shaped, then it cannot be ergodic because the RP cannot visit all the regions of $\Sigma_{\langle v\rangle (T),N}$ for a time proportional to their statistical measure. In other words, the neck of the $\Sigma_{\langle v\rangle (T),N}$ suppresses exponentially with $N$ the probability of the RP to jump form a lobe to the other, so that the neck acts similarly to a topological barrier. We note that, despite we have applied this idea to the canonical ensemble, this mechanism is suitable for application also to the microcanonical ensemble because of its independence on the thermodynamic limit.  

Even though not developed in this paper, the concept of dumbbell-shaped $\Sigma_{v,N}$'s can be directly extended also to discrete-variables systems, e.g., the Ising model. $\Sigma_{v,N}$'s become discrete sets, so that the statistical measure is replaced by the count of the microstates. 

The question weather a narrow neck may break the symmetry of a system was pointed out also in Refs. \cite{gfp,gfp1}, where the authors introduced the concept of "asymptotic diffeomorphicity" among manifolds: two manifolds can be diffeomorphic at any finite $N$, but not be asymptotically diffeomorphic in the limit $N\rightarrow\infty$. We wonder weather this picture may be equivalent to the one of dumbbell-shaped $\Sigma_{v,N}$'s put forward in this paper for systems with a $\mathbb{Z}_2$ symmetry for which dumbbell-shaped $\Sigma_{v,N}$'s are defined. 

Another open question is about short-range systems. In such systems the entropy $s(v,m)$ has to be a concave $v$-function, at most non-strictly concave in presence of a phase transition \cite{g,l}. The definition of dumbbell-shaped $\Sigma_{v,N}$'s we have given in this paper is suitable only for long-range systems because it implies a non-concave shape of $s(v,m)$. Anyway, that definition is suitable to be generalized to the short-range case if we assume that the $\Sigma_{v,N}$'s are dumbbell-shaped for any $N$ but not in the limit $N\rightarrow\infty$. This entails that the entropy must be a non-strictly concave $v$-function, as requested for a short-range system. This may means that there is no qualitative difference between short-range and long-range systems as long as $N$ is far from the thermodynamic limit. 

Lastly, a future research line may be try to extend the framework introduced in this paper for a $\mathbb{Z}_2$ symmetry to other symmetry groups, e.g., $O(n)$ with $n>1$.

\begin{acknowledgments}
	I warmly thank Matteo Gori for very useful discussions. 
\end{acknowledgments}

\end{document}